\documentclass[manuscript,screen]{acmart}
\AtBeginDocument{%
  \providecommand\BibTeX{{%
    \normalfont B\kern-0.5em{\scshape i\kern-0.25em b}\kern-0.8em\TeX}}}

\setcopyright{rightsretained}
\copyrightyear{2024}
\acmYear{2024}
\acmDOI{XXXXXXX.XXXXXXX}

\acmConference[TREW@CHI'24]{Make sure to enter the correct
  conference title from your rights confirmation emai}{May 11,
  2024}{Honolulu, HI}
\acmBooktitle{TREW@CHI'24: Workshop on Trust and Reliance in Evolving Human-AI Workflows,
 May 11, 2024, Honolulu, HI} 
\acmISBN{978-1-4503-XXXX-X/18/06}

\usepackage{soul}

\begin{document}

\title{Could Humans Outshine AI in Visual Data Analysis?}


\author{Ratanond Koonchanok}

\email{rkoonch@iu.edu}
\orcid{0000-0002-8860-6183}

\affiliation{%
  \institution{Indiana University Indianapolis}
  \streetaddress{}
  \city{Indianapolis}
  \state{Indiana}
  \country{USA}
  \postcode{46202}
}

\author{Khairi Reda}

\email{redak@iu.edu}
\orcid{0000-0002-8096-658X}

\affiliation{%
  \institution{Indiana University Indianapolis}
  \streetaddress{}
  \city{Indianapolis}
  \state{Indiana}
  \country{USA}
  \postcode{46202}
}

\renewcommand{\shortauthors}{Koonchanok and Reda}

\begin{abstract}

People often use visualizations not only to explore a dataset but also to draw generalizable conclusions about underlying models or phenomena. While previous research has viewed deviations from rational analysis as problematic, we hypothesize that human reliance on non-normative heuristics may be advantageous in certain situations. In this study, we investigate scenarios where human intuition might outperform idealized statistical rationality. Our experiment assesses participants' accuracy in characterizing the parameters of known data-generating models from bivariate visualizations. Our findings show that, while participants generally demonstrated lower accuracy than statistical models, they often outperformed Bayesian agents, particularly when dealing with extreme samples. These results suggest that, even when deviating from rationality, human gut reactions to visualizations can provide an advantage. Our findings offer insights into how analyst intuition and statistical models can be integrated to improve inference and decision-making, with important implications for the design of visual analytics tools.







\end{abstract}




\maketitle

\section{Introduction}

Visualization tools play an increasingly important role in data analysis and decision-making. With recent advances in LLM applications, people can quickly generate visualizations even without having an analytical background. One significant impact of those advanced AI tools such as ChatGPT on data visualization is their ability to enable conversational queries and commands. Users can interact with visualizations using plain language, asking questions, or requesting specific visualizations. There has also been a growing number of proposals for visualization-specific tools that utilize LLMs as their core component \cite{maddigan2023chat2vis, wang2023llm4vis, li2024prompt4vis}. While such advanced tools provide assistance in seeking insights from data \cite{zhao2024leva, wang2024scientific}, making accurate inferences from visualizations can still be challenging, as it requires one to carefully account for potential uncertainties and variabilities in the data. For example, the analyst might overinterpret a visualization displaying an unusual or outlying sample, causing them to infer spurious features that do not exist in the data-generating process (i.e., false discovery). Conversely, they may fail to sufficiently consider data that is in front of them, and instead fall back to their prior belief --- a potential manifestation of confirmation bias. 

We are motivated by the idea of collaborative exploration and interpretation of visualizations, where human cognition works alongside AI systems to derive insights from complex datasets. Human analysts often bring to bear their own hunches when interpreting visualizations. In coming to an inference, the analyst might choose to adaptively overweigh their prior information and underweigh the evidence implied by a dataset, particularly if the latter is perceived to be noisy or unreliable~\cite{lin2022data}. Such intuitive thinking is often regarded as untrustworthy and problematic~\cite{kahneman1982judgment, gilovich2002heuristics}. However, recent research has highlighted the role of heuristics in promoting accurate decisions~\cite{albrechtsen2009can, gigerenzer2008heuristics}. Intuitive decision-making can sometimes lead to better outcomes than rational, analytical reasoning~\cite{dane2012should, sadler2004intuitive}, particularly in high-risk high-uncertainty situations~\cite{huang2018role}. Non-statistically rational thinking could, thus, be useful when viewing a visualization with high uncertainty or `extreme' data. In essence, human inference-making can serve as a valuable complement to a strictly rational agent. 

In this study, we investigate situations where human visual inferences might outperform those of an ideal AI machine in accurately characterizing data-generating processes under varying uncertainty and sample conditions. Rather than assuming that inferences from an AI are always ideal, we compare the benefits and drawbacks of both sides, evaluating conditions in which the strength of one can compensate for the weakness of the other. To achieve this objective, we employ Bayesian models to assess how both humans and AI machines make inferences on datasets. Bayesian models offer a principled framework for incorporating prior beliefs and updating them based on observed evidence, making them suitable for analyzing complex datasets and assessing decision-making processes. Previous research has explored the application of Bayesian cognition approaches to data interpretation through visualization
\cite{kim2019bayesian, kim2020bayesian} where they evaluated the effectiveness of visualization techniques based on the alignment between human interpretations and the Bayesian model's conclusions. While this method can be helpful in assessing overall effectiveness, relying solely on a model as a reliable reference may not always be ideal. In certain contexts, deviations from AI's rationality can be advantageous. For instance, a model might perceive a minor positive correlation between two unrelated variables, such as the number of ice cream cones sold and the number of shark attacks. In contrast, a human analyst might become more skeptical upon observing the same data, irrationally reinforcing their belief in no correlation, eventually avoiding a false-positive conclusion. This perspective acknowledges that human heuristics, though often simplistic and seemingly inferior~\cite{tversky1974judgment}, can lead to more accurate, ecological inferences~\cite{gigerenzer2009homo}. 

By characterizing factors that lead analysts to exceed machine-inference performance, we lay the foundation for designing human-in-the-loop systems capable of leveraging both the intuition of analysts and the computational power of machines to harness insights from visualizations. These systems could integrate analyst intuition, such as through natural language queries, into their decision-making processes, enhancing their ability to provide meaningful insights and recommendations. To explore the opportunities, we ask the following research questions.

\textbf{RQ1: } How do human visual analysts compare to Bayesian agents that see that same data? We specifically compare the visual inferences of analysts to Bayesians: one informed with the prior knowledge of the human analyst, and another with a uniform prior that provides no preexisting knowledge before exposure to the data. 

\textbf{RQ2: } How do characteristics of data samples affect visual inference accuracy relative to Bayesian agents?   

We address these questions empirically in two crowdsourced studies. Specifically, we assess how accurately individuals perceive the true bi-variate relationship between pairs of attributes, upon being exposed to (potentially noisy) samples. We collect responses from viewers both before and after exposure to visualizations, allowing us to capture both their prior beliefs and posterior inferences about the data-generating process. We then measure the accuracy of human-vs-Bayesian inferences under different sample configurations, ranging from large samples that closely conform to the true parameters to small and potentially extreme samples. The work presented in this paper is a preliminary version of the research described in Koonchanok et al.~\cite{koonchanok2024trust}, where we provide a more detailed analysis and additional experiments.

\section{Methods}

We address our research questions empirically in a crowdsourced study where we assess how accurately individuals perceive the true bi-variate relationship between pairs of attributes when exposed to (potentially noisy) samples. We collect responses from viewers both before and after exposure to visualizations, allowing us to capture both their prior beliefs and post-sample (posterior) inferences about the data-generating process. We then measure the accuracy of human-vs-Bayesian inferences under different sample extremeness, ranging from samples that closely conform to the true parameter to the extreme, potentially misleading samples that arise sporadically. 

\begin{figure*}[]
\center
\includegraphics[width=0.9\linewidth]
{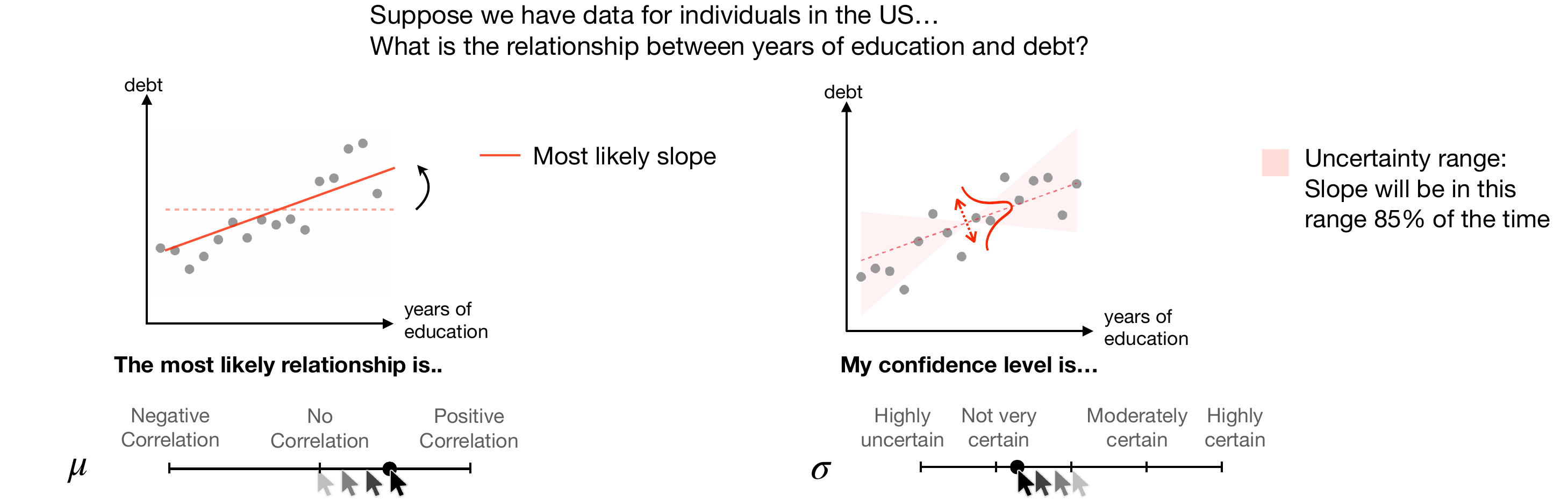}
  \caption{Elicitation Method. Participants utilize the first slider to adjust the correlation line (left). They then use the second slider to adjust the uncertainty range (right). Data points on the plot are refreshed at the frequency of 5 Hz to allow participants to see the model implication. }
  \label{fig:elicitation}
\end{figure*}

\subsection{Prior and Posterior Inference Elicitation}
\label{sec:elicitation}

We capture participants' prior and posterior beliefs through a graphical inference method, as illustrated in Figure \ref{fig:elicitation}. Previous research has shown that eliciting priors can be beneficial in multiple aspects \cite{koonchanok2021data, koonchanok2023visual, kim2017explaining, choi2019concept, heyer2020pushing, choi2019visual}. This interface enables participants to articulate two key parameters regarding their beliefs: the most likely true correlation coefficient between the two variables ($\mu$) and the associated uncertainty ($\sigma$). Participants enter their (prior or posterior) beliefs about these two parameters using two sliders. The first prompts them to indicate the ``most likely relationship'' along a continuum from `negative' to `positive' correlation. The second slider prompts participants to express their ``confidence level'' in the relationship, ranging from `highly uncertain' to moderately uncertain to `highly certain'. These two parameters are then used to populate the following model: 

\begin{equation} 
\begin{aligned}
    y_i = \beta_0 + \beta  x_i + \epsilon_i \\ \label{eq:model}
    \beta \sim \mathcal{N}(\mu, \sigma^2)\\
    \beta_0 \sim \mathcal{N}(0, \sigma_b^2) \\
    \epsilon_i \sim \mathcal{N}(0, {\sigma_e^2})
\end{aligned}
\end{equation}

Where $\mu \in (-1, 1)$ is the \emph{expected slope} of the relationship as specified by the relationship slider, and $\sigma \in (0, 0.6)$ is the \emph{uncertainty} in the slope, specified in the second confidence slider. $\beta_0$, an intercept for the regression line, centered around 0 with a fixed standard deviation of $\sigma_b=0.1$. $\sigma_e$ is a standard deviation of an additional residual term ($\epsilon_i$)  and is fixed at 0.45. We also display a hypothetical outcome plot (HOP), during which a new set of data points is refreshed every 200 milliseconds. 

\subsection{Stimuli and Data-Generating Models}
\label{sec:questions}

\begin{table*}
    \centering
    \begin{tabular}{c|l|c} 
         \textbf{Correlation} &  \textbf{Question} & \textbf{Ground Truth} \\ \hline \hline 
          Positive&  What is the relationship between teachers' experience & $\mu= 0.502$\\
 & and the average standardized test score for students?&$\sigma=  0.172$\\ \hline 
          Negative&  What is the relationship between time spent on social & $\mu= -0.422$\\
 & media and hours of sleep at night?&$\sigma=0.146 $\\ \hline
 No relationship& What is the relationship between the length of people's &$\mu= 0.030$\\
 & first name and last name?&$\sigma=0.236 $\\ 
    \end{tabular}
    \caption{Examples from the prompt questions used in our experiments. We collected a mean crowdsourced ground truth of the $\mu$ to determine both the correlation of and consensus level of each question.}
    \label{tab:example_questions}
\end{table*}

To provide plausible stimuli for the study, we developed a set of 40 prompt questions and corresponding ground truth models. The prompt questions covered a variety of common knowledge topics (see Table~\ref{tab:example_questions} for examples), with bivariate relationships ranging from negative correlations to positive correlations. Additionally, we also included attributes with no plausible relationship. The corresponding ground truth models for these prompts followed the same form as Equation~\ref{eq:model}. To initialize these models with plausible parameters rooted in common wisdom, we employed crowd workers recruited through Amazon Mechanical Turk. Workers ($n=61$) were prompted to respond to each of the 40 questions, using the model elicitation interface in \ref{fig:elicitation} to provide their belief on the most likely relationship slope and their uncertainty around that relationship.  Responses from the workers were then averaged forming the two ground truth parameters for each question. We then selected 24 prompt questions, based on the topic variety, to be used in the experiment. These questions comprised six ground truth models with a positive relationship ($\mu>0$), six questions with a negative relationship ($\mu<0$), and 12 with no correlation ($\mu\approx0$).  

\subsubsection{Social consensus}

 Individuals often rely on social knowledge when forming their beliefs. Social consensus (whether perceived or actual) can also serve as a tool to reduce uncertainty. Therefore, we expect the agreement around the ground truth to impact people's inferences from visualizations. To quantify the latter, we measure the consistency of the crowd wisdom: Prompt questions exhibiting smaller variations in the elicited crowd beliefs are considered to reflect a higher degree of social consensus. This was determined based on the standard deviation of the elicited $\mu$ among the workers. Within each category (positive, negative, or no relationship), we designate half the question with the lowest standard deviation as `high' consensus, with the other half considered `low' consensus, representing lower agreement between workers on what the data-generating process should be.

\subsection{Sample Properties}
\label{sec:sampleProperties}

For each stimulus, we display the prompt question and ask the participant to provide their prior belief about the topic. The participant specifies their two prior knowledge parameters ($\mu$ and $\sigma$) using the belief elicitation device. We then expose participants to a random data sample generated from a ground truth model. Lastly, we ask them to provide a posterior inference using the same elicitation device as before. To understand how sample characteristics affect inference accuracy, we varied two sample properties: size and extremeness. 

\subsubsection{Sample Size}

The number of observations in a sample is an important factor for an analyst to consider when making inferences. Larger sample sizes generally supply stronger evidence about the underlying data-generating sample. We thus varied the size of samples shown to participants, using 7, 15, and 30 data points to represent `small', `medium', and `large' sample sizes, respectively. These sample sizes were selected to allow for varying levels of evidence, while still allowing for extreme samples to emerge. Specifically, smaller samples are more prone to noise, giving a potentially misleading picture of the underlying ground truth. 

\subsubsection{Sample Extremeness}

A core question that we address in this work is how resilient people are to extreme (or spurious) samples, as compared with idealized statistical machines. Sample extremeness reflects the degree to which it is consistent with the ground truth model. Under a random sampling regime, spurious samples are expected to naturally arise in the study. Consequently, we do not explicitly manipulate sample extremeness. Instead, we simply quantify and record the extremeness of generated stimuli in every trial. Under the law of large numbers, we expect samples generated to conform to the ground truth on the aggregate. In particular, the majority of samples should reflect the expected correlation coefficient of the corresponding ground truth model. However, extreme and spurious would be expected to emerge in the experiment (e.g., a sample showing a positive correlation when the underlying ground truth indicates no relationship).

We use Pearson's coefficient ($R$) to characterize the strength of correlation for a sample and, by extension, the degree to which it can be considered \emph{extreme} with respect to its underlying ground truth model. Specifically, we compute a difference ($\Delta R$) between the sample's correlation and the expected strength of the correlation:

\begin{equation}
\begin{aligned}
\Delta R = R_{Sample} - R_{Expected} \label{eq:extremeness}
\end{aligned}
\end{equation}

Where $R_{Sample}$ is the sample's correlation coefficient and  $R_{Expected}$ is the mean R-value, determined from 10,000 simulated draws from the ground truth model. A sample that is a faithful representation of the data-generating model will have $\Delta R \approx 0$. As $\Delta R$ deviates (either negatively or positively), the sample will be considered more extreme. Under the law of large numbers, we would expect the distribution of $\Delta R$ to be centered around zero, with the majority of samples exhibiting low extremeness. Furthermore, we would expect the sample size to affect the likelihood of extreme samples: a sample with a larger number of data points will exhibit a narrower $\Delta R$ distribution. Conversely, samples with fewer data points will exhibit a wider extremeness distribution. Data-generating models with high uncertainty ($\sigma$) will also exhibit a greater likelihood of extreme samples that deviate from the expected slope. 

By measuring sample extremeness during the study, we can observe participants' reactions to spurious samples when those arise. Furthermore, we can gauge the degree to which participants (and Bayesian agents) are resilient to those misleading samples, and whether either can recover the underlying ground truth parameters. 

\section{Experiment}

The experiment aims to compare the inference accuracy between participants to ideal Bayesian agents. We specifically test the extent to which participants can infer the ground truth parameters as a function of sample size, sample extremeness, and the degree of social consensus (along with interactions of these factors). Participants saw prompt questions described in \S\ref{sec:questions}, and were asked to predict the parameters of a bi-variate relationship model, namely, the expected slope $\mu$ and uncertainty $\sigma$. In each stimulus, participants were asked to provide their prior beliefs through the graphical elicitation interface described earlier. Specifically, participants adjusted the two sliders in Figure~\ref{fig:elicitation}, until the observed HOP sample matched their expectation of the relationship. Following the prior specification, they were then shown a data sample drawn randomly from the ground truth model side-by-side with their prior belief. During this step, they were asked to indicate how reliable they believed the sample to be, providing their subjective rating via a slider. Upon inspecting the sample, participants were prompted to provide their updated (posterior) belief, which reflect their inference ``about the true relationship'' via the same elicitation device. We specifically instructed participants to re-adjust the two parameter sliders again after considering the sample. The entire steps are shown in Figure \ref{fig:steps}.

\subsection{Experiment Design}

We adopt a mixed design, investigating two independent factors (varied within subjects): Sample size (3 levels) $\times$ Social consensus (2 levels). 

\noindent \textbf{Sample characteristics}: We varied sample characteristics within-subject. Specifically, participants completed an equal number of 8 trials with each sample size (small, medium, large). Participants also saw an equal number of low and high-consensus prompts. Sample extremeness was included in our analysis as an explanatory variable. Extremeness was left to vary as a consequence of the random sampling process. Hence, while not controlled on a participant level, extremeness followed a predictable distribution and was used as one of the explanatory factors in our modeling and analysis of the results (see \S\ref{sec:sampleProperties} for a discussion of this).

\noindent \textbf{Inferences}: Participants supplied four continuous responses in every trial: the expected slope ($\mu$) and uncertainty ($\sigma$) in the relationship, both before (i.e., prior belief) \emph{and} after sample exposure (i.e., posterior inference). For every trial, we generated two comparable posterior inferences. The first ($\mu_{Bayes}$ and $\sigma_{Bayes}$) was derived through Bayesian inference using the participant-provided prior and the likelihood function implied by the same sample observed by the participant. This corresponds to an idealized agent equipped with identical prior knowledge as the participant. The second response, also a Bayesian agent, utilized a flat prior in conjunction with the same sample-implied likelihood, thus representing a `blank-slate' agent that exclusively learns the two parameters ($\mu_{Flat}$ and $\sigma_{Flat}$) purely from the sample. We measured the distance between the ground truth parameters (i.e., true $\mu$ and $\sigma$) and the posterior separately for each of the three agents (Participant, Bayesian, and Flat Bayesian). 

\subsection{Participants}
Potential participants were recruited through Prolific. We ended up recruiting 74 participants. We recruited workers who are US residents with a minimum task-approval rate of 98\%. Participants received a \$5 compensation upon completing the experiment. The study was approved by Indiana University's institutional review board.

\begin{figure*}[]
\center
\includegraphics[width=1\linewidth]
{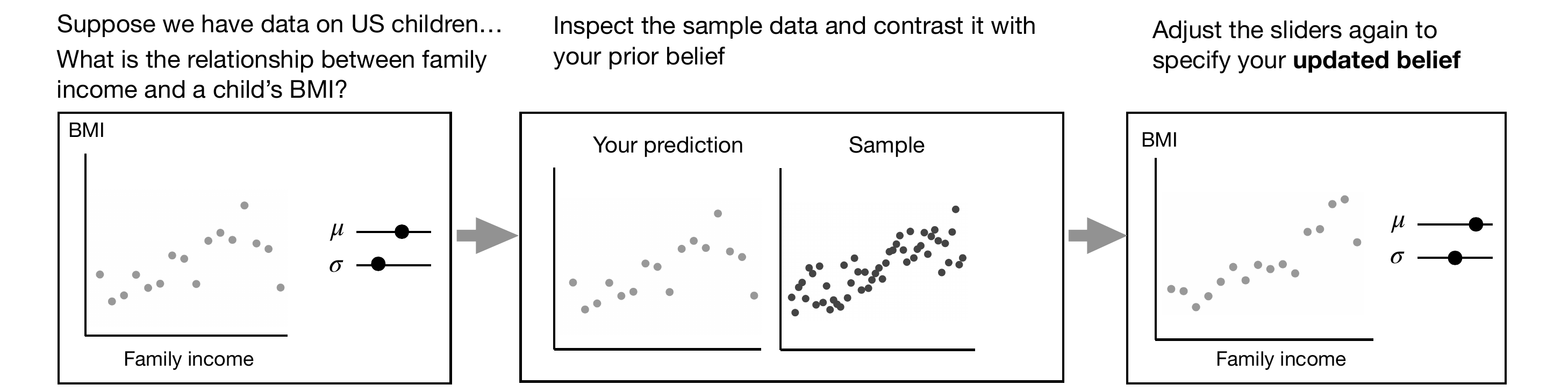}
  \caption{Steps per trial that each participant. participants start by specifying initial belie of $\mu$ and $\sigma$ using the two sliders. They then observe the data sample and compare it against their prediction. Subsequently, they specify their posterior beliefs using the same interface as the first step.}
  \label{fig:steps}
\end{figure*}

\subsection{Procedure}

Before starting the experiment, participants went through a series of tutorials explaining the goal of the study and explained how the graphical elicitation device worked. The instructions emphasized that the samples shown could be noisy, particularly when the sample size is small. Participants then completed a practice trial. After the tutorial, participants completed the main study which consisted of 24 trials, corresponding to the prompt questions developed in \S\ref{sec:questions}. In addition to the analyzed trails, we included four engagement checks inserted randomly within the trial stream. The engagement questions consisted of questions with the same setup as the rest but with different instructions asking participants to move sliders to a specific location (e.g. extreme right). 

In each trial, participants were first presented with a context (e.g., supposed we have data on US cities), and were then prompted to predict a linear relationship between two variables (e.g. what is the relationship between the unemployment rate and rate of affordable housing?) The participant was asked to visually provide their prior belief through the elicitation device described in \S\ref{sec:elicitation}. Next, the participants were presented with a sample drawn from the ground-truth model, shown side-by-side with their prior. Lastly, the participant was prompted to update their prior belief by re-adjusting the same set of two sliders. Figure~\ref{fig:steps} illustrates the sequence of these steps for a trial. 

\subsection{Accuracy Metrics}

To characterize the accuracy of posterior inferences, we a metric that measures the divergence between the posterior and the true values. For readability, we normalize to obtain relative maximum/minimum deviation based on the range of parameters dictated by the sliders ([-1, 1]). Specifically:

\begin{equation}
\begin{aligned} 
   \Delta \mu &= \frac{ \mu_{Human|Bayes|Flat} - \mu_{Truth} }{2}   \\
\end{aligned}
\end{equation}

 The \emph{Human} response is the posterior inference of the participant, whereas \emph{Bayes} and \emph{Flat} are the normative Bayesian inferences that are either based on the participant-supplied prior or an assumption of a flat prior, respectively. In the subsequent results, we multiplied the number by 100 to represent percentage values. 
 
\section{Results}

We used the brms package \cite{burkner2017brms} to fit the responses to a Bayesian regression model. The model predicts  $\Delta \mu$. We modeled both the \emph{mean} and \emph{variance} of the divergence, although the mean is our primary interest. We started by constructing a simpler model formulation that included the primary experimental factors. We then followed a model-expansion approach, adding interaction effects to improve the model fit (assessed using posterior predictive simulations). We also added random effects to account for individual differences among participants. 

\begin{equation*}
\begin{aligned} 
    \Delta \mu  &\sim Normal(\text{mean}, \text{sd}^2) \\
    \text{mean} &= \Delta R \times agent \times size \times consensus  
         + \Delta R\times agent \times questionType  \\
         &+ (1+\Delta R \times agent ~|~participant) 
         + (1~|~question) \\
    log(\text{sd}) &= agent \times size \times consensus + (1+size \times agent~|~participant) + (1|question) \\
\end{aligned}
\end{equation*}

$\Delta R$ is the sample extremeness (as defined in Equation~\ref{eq:extremeness}), \emph{agent} indicates whether the inference came from a Human (the participant), a Bayesian, or flat-prior agent. \emph{Size} is a categorical variable representing the sample size (Small, Medium, or Large), \emph{consensus} is the social consensus around the ground truth, and \emph{questionType} is a categorical variable indicating whether the true model prescribes positive correlation, negative correlation, or no relationship. 

Participants completed the experiment in 16.39 minutes on average. They provided 1776 responses in total. We first compare how participants perform compared with the Bayesian machines. We then analyze how each experimental factors affect the inference accuracy.

\subsection{Human vs machine inferences}

\begin{figure}[]
\center
\includegraphics[width=0.5\linewidth]
{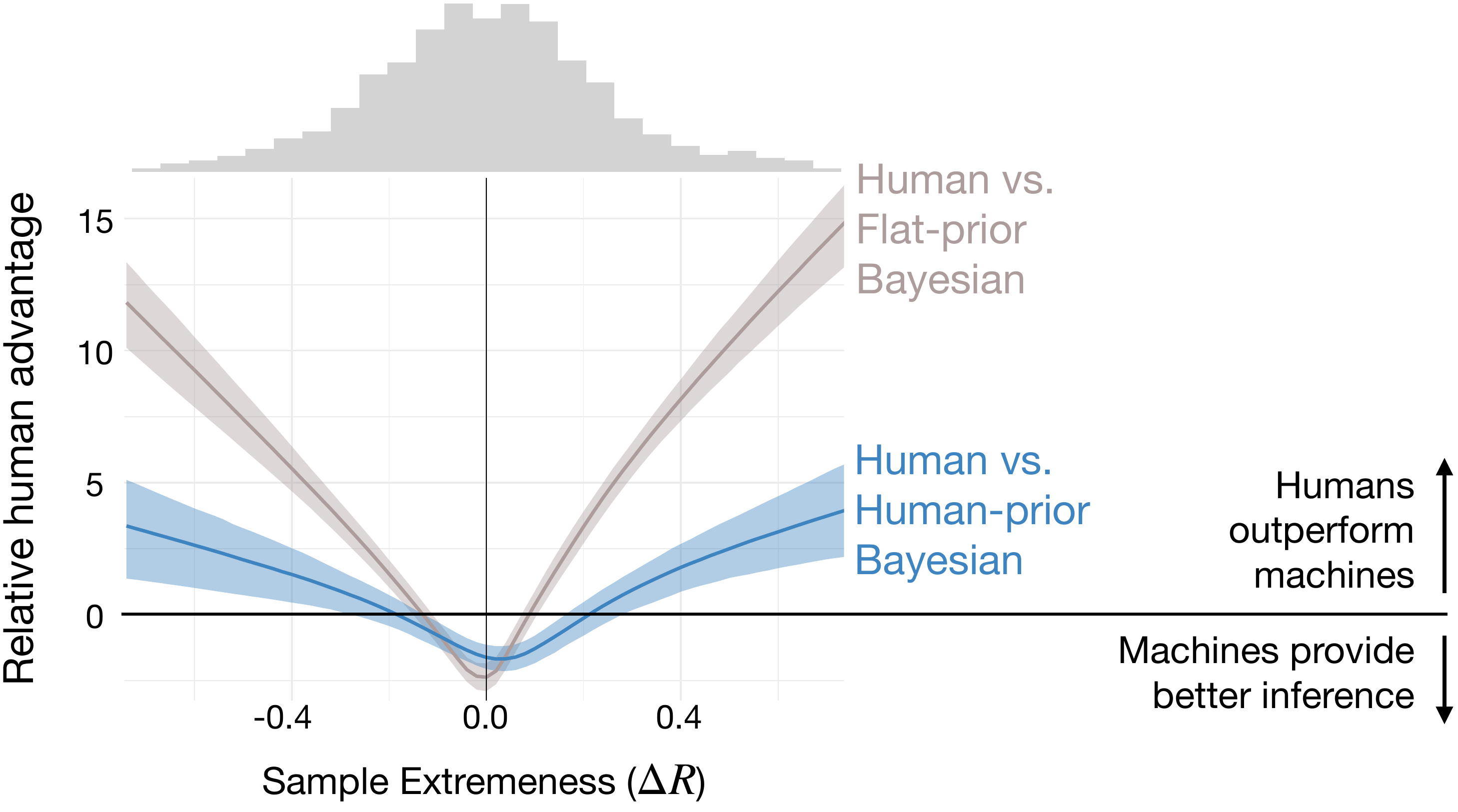}
  \caption{Comparison between humans and statistical machines in inferring true $\mu$, contingent on sample extremeness. The top histogram illustrates the empirical distribution observed at various $\Delta R$ levels.}
  \label{fig:exp1_HumanMachine}
\end{figure}

Figure~\ref{fig:exp1_HumanMachine} illustrates the relative resilience of humans compared to statistical inference. Specifically, we depict the estimated difference in divergence between participants on one hand and informed Bayesian agents (blue) and uninformed Bayesian agents (grey) on the other.
An estimate above 0 suggests superior inferential performance for humans compared to statistical machines, while a value below zero indicates the opposite. When $\Delta R=0$ (indicating a sample fully consistent with the ground truth), the model predicts that humans will underperform, showing a deficit of $-1.62$ (CI: [$-2.06, -1.14$]) against informed Bayesian agents and $-2.37$ (CI: [$-2.90, -1.84$]) relative to an uninformed, data-driven agent. At $\Delta R = 0.2$ , however, the model predicts nearly equal performance for humans and informed Bayesian agents and a substantial advantage for humans over an uninformed agent. The performance advantage for humans can be expected to further widen at $\Delta R = .4$, relative to informed Bayesian.

\subsection{To what extent are humans  resilient to misleading samples compared to Bayesian agents?}

\begin{figure}[]
\center
\includegraphics[width=1
\linewidth]{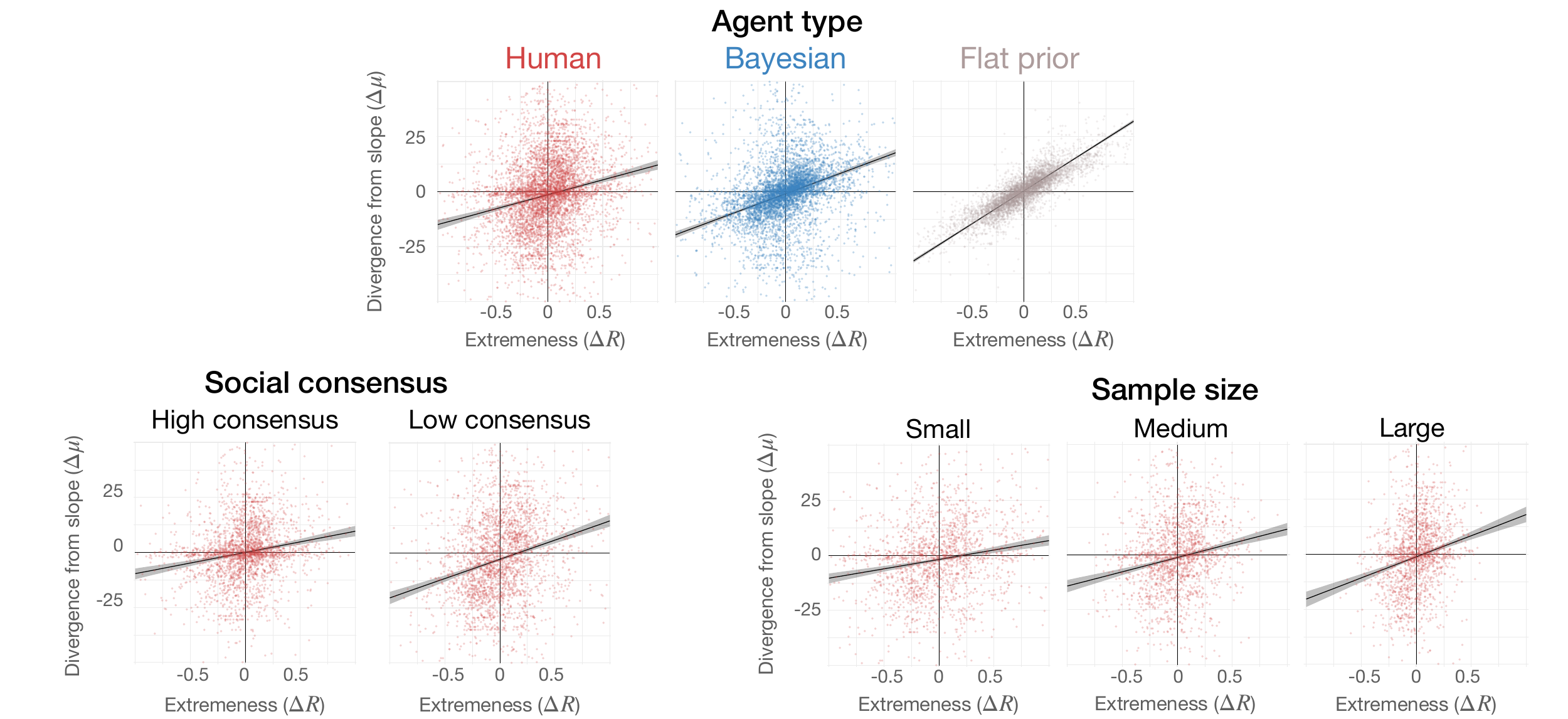}
  \caption{The sensitivity of the three agents to sample extremeness. The regression line shows model estimates. Points depict the observed, empirical responses. A weaker correlation implies more resilience to spurious samples.}
  \label{fig:exp1_lineCharts}
\end{figure}

Figure \ref{fig:exp1_lineCharts}-top illustrates the impact of sample extremeness on the divergence from the ground truth for the three agent types. A weaker correlation (i.e., smaller slope) indicates lower sensitivity and, hence, better resilience to spurious samples. Participants appear to be less influenced by extreme samples than both informed and uninformed (flat prior) Bayesian agents. Specifically, a unit increase in sample extremeness leads to a 13.53 (95\% CI: [$11.43, 15.82$]) increase in divergence from the true relationship. By contrast, a Bayesian agent is influenced more strongly by misleading samples (18.61, CI: [$17.1, 20.16$]), leading to higher divergence at more extreme samples. As would be expected, an agent with a flat prior learning purely from the data is influenced the most (31.73, CI: [$30.96, 32.52$]), leading to a strong association between sample quality and inference accuracy. When there is more consensus around the ground truth, participants are more resilient to spurious samples (9.65, CI: [7.32, 12.11] for high consensus vs., 17.46 CI: [14.82, 20.29] for low consensus). Lastly, sample size also plays a role, with participants most skeptical of small samples (8.57, CI: [6.36, 10.91]), followed by medium (12.90, CI: [10.16, 15.74]) and large (19.21, CI: [15.79, 22.77]). For both Bayesian agents, their sensitivities to extremeness by social consensus and sample size closely align with the overall estimate.

\section{Discussion}

Overall, both informative and uninformative Bayesian machines are more accurate than participants when data samples closely represent the data-generating models. However, as samples become more extreme, participants become more accurate relative to the Bayesian machines. Participants eventually outperform the machines at certain points and continue to widen the gap as extremeness increases. Even though participants outperform the Bayesian machines for a wide range of extremeness, the machines are still more accurate overall because extreme samples are less likely to be generated.

Our results reveal that individuals often deviate from normative inference while analyzing visualizations. Interestingly, these deviations can be beneficial, especially when viewing visualizations that potentially include extreme data from small sample sizes. In such cases, an analyst's intuition may afford more accurate inferences than those produced by an ideal inference machine, even if the machine is calibrated with the participants' pre-existing knowledge. Conversely, for larger and more reliable datasets, the precision offered by machine-based inferences can reduce bias, leading to more accurate inference. 

These findings suggest that humans and machines excel in different scenarios, highlighting the potential for collaboration between humans and AI-based agents in inference-making. This collaboration allows each to complement the limitations of the other~\cite{wang2020human, steyvers2022bayesian}. For instance, a system could offer feedback and recommendations on visual inference, taking into account the sample's characteristics and the user's prior knowledge.  If the current sample seems reliable, factoring in aspects like sample size, then one might lean toward relying on a machine. Conversely, in cases where the data appears noisy, relying on human heuristics might prove more suitable. While including humans as part of a system has demonstrated benefits in several machine learning studies \cite{daee2018user, mozannar2022teaching, steyvers2022bayesian, wang2022interpretability, salimzadeh2023missing}, the challenge in building such visual analytics systems is to determine how much trust to allocate to the AI and when to be skeptical. One reason is that it is not straightforward to pre-determine factors such as sample extremeness when encountering new data. Furthermore, it might be a challenge for humans to accurately elicit prior knowledge that reflects statistical parameters. With the assistance of LLMs, we believe that the latter challenge could be addressed. Through natural language input, we have the potential to enhance precise knowledge elicitation, thereby bolstering the effectiveness of visual data analysis. 

\section{Conclusion}

In this study, we investigate whether human inference can outperform AI models when making inferences from graphical patterns. We conducted a crowdsourced study to compare the inference accuracy of humans against Bayesian agents, incorporating multiple factors such as sample size and sample extremeness. Participants externalized both their prior beliefs (i.e., before seeing data) and posterior beliefs (i.e., after data exposure) using graphical elicitation. Our results demonstrate that humans were more accurate at inferring the true correlation level when the visualization depicted extreme data samples. These results underscore the potential effectiveness of integrating human-AI workflows to enhance the quality of decision-making.

\section*{Acknowledgement}
This paper is based upon research supported by the National Science Foundation under award 1942429.

\bibliographystyle{ACM-Reference-Format}
\bibliography{00_ref}


\begin{thebibliography}{31}


\ifx \showCODEN    \undefined \def \showCODEN     #1{\unskip}     \fi
\ifx \showDOI      \undefined \def \showDOI       #1{#1}\fi
\ifx \showISBNx    \undefined \def \showISBNx     #1{\unskip}     \fi
\ifx \showISBNxiii \undefined \def \showISBNxiii  #1{\unskip}     \fi
\ifx \showISSN     \undefined \def \showISSN      #1{\unskip}     \fi
\ifx \showLCCN     \undefined \def \showLCCN      #1{\unskip}     \fi
\ifx \shownote     \undefined \def \shownote      #1{#1}          \fi
\ifx \showarticletitle \undefined \def \showarticletitle #1{#1}   \fi
\ifx \showURL      \undefined \def \showURL       {\relax}        \fi
\providecommand\bibfield[2]{#2}
\providecommand\bibinfo[2]{#2}
\providecommand\natexlab[1]{#1}
\providecommand\showeprint[2][]{arXiv:#2}

\bibitem[Albrechtsen et~al\mbox{.}(2009)]%
        {albrechtsen2009can}
\bibfield{author}{\bibinfo{person}{Justin~S Albrechtsen}, \bibinfo{person}{Christian~A Meissner}, {and} \bibinfo{person}{Kyle~J Susa}.} \bibinfo{year}{2009}\natexlab{}.
\newblock \showarticletitle{Can intuition improve deception detection performance?}
\newblock \bibinfo{journal}{\emph{Journal of Experimental Social Psychology}} \bibinfo{volume}{45}, \bibinfo{number}{4} (\bibinfo{year}{2009}), \bibinfo{pages}{1052--1055}.
\newblock


\bibitem[B{\"u}rkner(2017)]%
        {burkner2017brms}
\bibfield{author}{\bibinfo{person}{Paul-Christian B{\"u}rkner}.} \bibinfo{year}{2017}\natexlab{}.
\newblock \showarticletitle{brms: An R package for Bayesian multilevel models using Stan}.
\newblock \bibinfo{journal}{\emph{Journal of statistical software}}  \bibinfo{volume}{80} (\bibinfo{year}{2017}), \bibinfo{pages}{1--28}.
\newblock


\bibitem[Choi et~al\mbox{.}(2019a)]%
        {choi2019concept}
\bibfield{author}{\bibinfo{person}{In~Kwon Choi}, \bibinfo{person}{Taylor Childers}, \bibinfo{person}{Nirmal~Kumar Raveendranath}, \bibinfo{person}{Swati Mishra}, \bibinfo{person}{Kyle Harris}, {and} \bibinfo{person}{Khairi Reda}.} \bibinfo{year}{2019}\natexlab{a}.
\newblock \showarticletitle{Concept-driven visual analytics: an exploratory study of model-and hypothesis-based reasoning with visualizations}. In \bibinfo{booktitle}{\emph{Proceedings of the 2019 chi conference on human factors in computing systems}}. \bibinfo{pages}{1--14}.
\newblock


\bibitem[Choi et~al\mbox{.}(2019b)]%
        {choi2019visual}
\bibfield{author}{\bibinfo{person}{In~Kwon Choi}, \bibinfo{person}{Nirmal~Kumar Raveendranath}, \bibinfo{person}{Jared Westerfield}, {and} \bibinfo{person}{Khairi Reda}.} \bibinfo{year}{2019}\natexlab{b}.
\newblock \showarticletitle{Visual (dis) confirmation: Validating models and hypotheses with visualizations}. In \bibinfo{booktitle}{\emph{2019 23rd International Conference in Information Visualization--Part II}}. IEEE, \bibinfo{pages}{116--121}.
\newblock


\bibitem[Daee et~al\mbox{.}(2018)]%
        {daee2018user}
\bibfield{author}{\bibinfo{person}{Pedram Daee}, \bibinfo{person}{Tomi Peltola}, \bibinfo{person}{Aki Vehtari}, {and} \bibinfo{person}{Samuel Kaski}.} \bibinfo{year}{2018}\natexlab{}.
\newblock \showarticletitle{User modelling for avoiding overfitting in interactive knowledge elicitation for prediction}. In \bibinfo{booktitle}{\emph{23rd International Conference on Intelligent User Interfaces}}. \bibinfo{pages}{305--310}.
\newblock


\bibitem[Dane et~al\mbox{.}(2012)]%
        {dane2012should}
\bibfield{author}{\bibinfo{person}{Erik Dane}, \bibinfo{person}{Kevin~W Rockmann}, {and} \bibinfo{person}{Michael~G Pratt}.} \bibinfo{year}{2012}\natexlab{}.
\newblock \showarticletitle{When should I trust my gut? Linking domain expertise to intuitive decision-making effectiveness}.
\newblock \bibinfo{journal}{\emph{Organizational behavior and human decision processes}} \bibinfo{volume}{119}, \bibinfo{number}{2} (\bibinfo{year}{2012}), \bibinfo{pages}{187--194}.
\newblock


\bibitem[Gigerenzer(2008)]%
        {gigerenzer2008heuristics}
\bibfield{author}{\bibinfo{person}{Gerd Gigerenzer}.} \bibinfo{year}{2008}\natexlab{}.
\newblock \showarticletitle{Why heuristics work}.
\newblock \bibinfo{journal}{\emph{Perspectives on psychological science}} \bibinfo{volume}{3}, \bibinfo{number}{1} (\bibinfo{year}{2008}), \bibinfo{pages}{20--29}.
\newblock


\bibitem[Gigerenzer and Brighton(2009)]%
        {gigerenzer2009homo}
\bibfield{author}{\bibinfo{person}{Gerd Gigerenzer} {and} \bibinfo{person}{Henry Brighton}.} \bibinfo{year}{2009}\natexlab{}.
\newblock \showarticletitle{Homo heuristicus: Why biased minds make better inferences}.
\newblock \bibinfo{journal}{\emph{Topics in cognitive science}} \bibinfo{volume}{1}, \bibinfo{number}{1} (\bibinfo{year}{2009}), \bibinfo{pages}{107--143}.
\newblock


\bibitem[Gilovich et~al\mbox{.}(2002)]%
        {gilovich2002heuristics}
\bibfield{author}{\bibinfo{person}{Thomas Gilovich}, \bibinfo{person}{Dale Griffin}, {and} \bibinfo{person}{Daniel Kahneman}.} \bibinfo{year}{2002}\natexlab{}.
\newblock \bibinfo{booktitle}{\emph{Heuristics and biases: The psychology of intuitive judgment}}.
\newblock \bibinfo{publisher}{Cambridge university press}.
\newblock


\bibitem[Heyer et~al\mbox{.}(2020)]%
        {heyer2020pushing}
\bibfield{author}{\bibinfo{person}{Jeremy Heyer}, \bibinfo{person}{Nirmal~Kumar Raveendranath}, {and} \bibinfo{person}{Khairi Reda}.} \bibinfo{year}{2020}\natexlab{}.
\newblock \showarticletitle{Pushing the (visual) narrative: the effects of prior knowledge elicitation in provocative topics}. In \bibinfo{booktitle}{\emph{Proceedings of the 2020 CHI Conference on Human Factors in Computing Systems}}. \bibinfo{pages}{1--14}.
\newblock


\bibitem[Huang(2018)]%
        {huang2018role}
\bibfield{author}{\bibinfo{person}{Laura Huang}.} \bibinfo{year}{2018}\natexlab{}.
\newblock \showarticletitle{The role of investor gut feel in managing complexity and extreme risk}.
\newblock \bibinfo{journal}{\emph{Academy of Management Journal}} \bibinfo{volume}{61}, \bibinfo{number}{5} (\bibinfo{year}{2018}), \bibinfo{pages}{1821--1847}.
\newblock


\bibitem[Kahneman et~al\mbox{.}(1982)]%
        {kahneman1982judgment}
\bibfield{author}{\bibinfo{person}{Daniel Kahneman}, \bibinfo{person}{Paul Slovic}, {and} \bibinfo{person}{Amos Tversky}.} \bibinfo{year}{1982}\natexlab{}.
\newblock \bibinfo{booktitle}{\emph{Judgment under uncertainty: Heuristics and biases}}.
\newblock \bibinfo{publisher}{Cambridge university press}.
\newblock


\bibitem[Kim et~al\mbox{.}(2020)]%
        {kim2020bayesian}
\bibfield{author}{\bibinfo{person}{Yea-Seul Kim}, \bibinfo{person}{Paula Kayongo}, \bibinfo{person}{Madeleine Grunde-McLaughlin}, {and} \bibinfo{person}{Jessica Hullman}.} \bibinfo{year}{2020}\natexlab{}.
\newblock \showarticletitle{Bayesian-assisted inference from visualized data}.
\newblock \bibinfo{journal}{\emph{IEEE Transactions on Visualization and Computer Graphics}} \bibinfo{volume}{27}, \bibinfo{number}{2} (\bibinfo{year}{2020}), \bibinfo{pages}{989--999}.
\newblock


\bibitem[Kim et~al\mbox{.}(2017)]%
        {kim2017explaining}
\bibfield{author}{\bibinfo{person}{Yea-Seul Kim}, \bibinfo{person}{Katharina Reinecke}, {and} \bibinfo{person}{Jessica Hullman}.} \bibinfo{year}{2017}\natexlab{}.
\newblock \showarticletitle{Explaining the gap: Visualizing one's predictions improves recall and comprehension of data}. In \bibinfo{booktitle}{\emph{Proceedings of the 2017 CHI Conference on Human Factors in Computing Systems}}. \bibinfo{pages}{1375--1386}.
\newblock


\bibitem[Kim et~al\mbox{.}(2019)]%
        {kim2019bayesian}
\bibfield{author}{\bibinfo{person}{Yea-Seul Kim}, \bibinfo{person}{Logan~A Walls}, \bibinfo{person}{Peter Krafft}, {and} \bibinfo{person}{Jessica Hullman}.} \bibinfo{year}{2019}\natexlab{}.
\newblock \showarticletitle{A bayesian cognition approach to improve data visualization}. In \bibinfo{booktitle}{\emph{Proceedings of the 2019 chi conference on human factors in computing systems}}. \bibinfo{pages}{1--14}.
\newblock


\bibitem[Koonchanok et~al\mbox{.}(2021)]%
        {koonchanok2021data}
\bibfield{author}{\bibinfo{person}{Ratanond Koonchanok}, \bibinfo{person}{Parul Baser}, \bibinfo{person}{Abhinav Sikharam}, \bibinfo{person}{Nirmal~Kumar Raveendranath}, {and} \bibinfo{person}{Khairi Reda}.} \bibinfo{year}{2021}\natexlab{}.
\newblock \showarticletitle{Data prophecy: Exploring the effects of belief elicitation in visual analytics}. In \bibinfo{booktitle}{\emph{Proceedings of the 2021 CHI Conference on Human Factors in Computing Systems}}. \bibinfo{pages}{1--12}.
\newblock


\bibitem[Koonchanok et~al\mbox{.}(2024)]%
        {koonchanok2024trust}
\bibfield{author}{\bibinfo{person}{Ratanond Koonchanok}, \bibinfo{person}{Michael~E Papka}, {and} \bibinfo{person}{Khairi Reda}.} \bibinfo{year}{2024}\natexlab{}.
\newblock \showarticletitle{Trust Your Gut: Comparing Human and Machine Inference from Noisy Visualizations}.
\newblock \bibinfo{journal}{\emph{IEEE Transactions on Visualization and Computer Graphics}} (\bibinfo{year}{2024}).
\newblock


\bibitem[Koonchanok et~al\mbox{.}(2023)]%
        {koonchanok2023visual}
\bibfield{author}{\bibinfo{person}{Ratanond Koonchanok}, \bibinfo{person}{Gauri~Yatindra Tawde}, \bibinfo{person}{Gokul~Ragunandhan Narayanasamy}, \bibinfo{person}{Shalmali Walimbe}, {and} \bibinfo{person}{Khairi Reda}.} \bibinfo{year}{2023}\natexlab{}.
\newblock \showarticletitle{Visual Belief Elicitation Reduces the Incidence of False Discovery}. In \bibinfo{booktitle}{\emph{Proceedings of the 2023 CHI Conference on Human Factors in Computing Systems}}. \bibinfo{pages}{1--17}.
\newblock


\bibitem[Li et~al\mbox{.}(2024)]%
        {li2024prompt4vis}
\bibfield{author}{\bibinfo{person}{Shuaimin Li}, \bibinfo{person}{Xuanang Chen}, \bibinfo{person}{Yuanfeng Song}, \bibinfo{person}{Yunze Song}, {and} \bibinfo{person}{Chen Zhang}.} \bibinfo{year}{2024}\natexlab{}.
\newblock \showarticletitle{Prompt4Vis: Prompting Large Language Models with Example Mining and Schema Filtering for Tabular Data Visualization}.
\newblock \bibinfo{journal}{\emph{arXiv preprint arXiv:2402.07909}} (\bibinfo{year}{2024}).
\newblock


\bibitem[Lin et~al\mbox{.}(2022)]%
        {lin2022data}
\bibfield{author}{\bibinfo{person}{Haihan Lin}, \bibinfo{person}{Derya Akbaba}, \bibinfo{person}{Miriah Meyer}, {and} \bibinfo{person}{Alexander Lex}.} \bibinfo{year}{2022}\natexlab{}.
\newblock \showarticletitle{Data hunches: Incorporating personal knowledge into visualizations}.
\newblock \bibinfo{journal}{\emph{IEEE Transactions on Visualization and Computer Graphics}} \bibinfo{volume}{29}, \bibinfo{number}{1} (\bibinfo{year}{2022}), \bibinfo{pages}{504--514}.
\newblock


\bibitem[Maddigan and Susnjak(2023)]%
        {maddigan2023chat2vis}
\bibfield{author}{\bibinfo{person}{Paula Maddigan} {and} \bibinfo{person}{Teo Susnjak}.} \bibinfo{year}{2023}\natexlab{}.
\newblock \showarticletitle{Chat2vis: Generating data visualisations via natural language using chatgpt, codex and gpt-3 large language models}.
\newblock \bibinfo{journal}{\emph{IEEE Access}} (\bibinfo{year}{2023}).
\newblock


\bibitem[Mozannar et~al\mbox{.}(2022)]%
        {mozannar2022teaching}
\bibfield{author}{\bibinfo{person}{Hussein Mozannar}, \bibinfo{person}{Arvind Satyanarayan}, {and} \bibinfo{person}{David Sontag}.} \bibinfo{year}{2022}\natexlab{}.
\newblock \showarticletitle{Teaching humans when to defer to a classifier via exemplars}. In \bibinfo{booktitle}{\emph{Proceedings of the AAAI Conference on Artificial Intelligence}}, Vol.~\bibinfo{volume}{36}. \bibinfo{pages}{5323--5331}.
\newblock


\bibitem[Sadler-Smith and Shefy(2004)]%
        {sadler2004intuitive}
\bibfield{author}{\bibinfo{person}{Eugene Sadler-Smith} {and} \bibinfo{person}{Erella Shefy}.} \bibinfo{year}{2004}\natexlab{}.
\newblock \showarticletitle{The intuitive executive: Understanding and applying ‘gut feel’in decision-making}.
\newblock \bibinfo{journal}{\emph{Academy of Management Perspectives}} \bibinfo{volume}{18}, \bibinfo{number}{4} (\bibinfo{year}{2004}), \bibinfo{pages}{76--91}.
\newblock


\bibitem[Salimzadeh et~al\mbox{.}(2023)]%
        {salimzadeh2023missing}
\bibfield{author}{\bibinfo{person}{Sara Salimzadeh}, \bibinfo{person}{Gaole He}, {and} \bibinfo{person}{Ujwal Gadiraju}.} \bibinfo{year}{2023}\natexlab{}.
\newblock \showarticletitle{A Missing Piece in the Puzzle: Considering the Role of Task Complexity in Human-AI Decision Making}. In \bibinfo{booktitle}{\emph{Proceedings of the 31st ACM Conference on User Modeling, Adaptation and Personalization}}. \bibinfo{pages}{215--227}.
\newblock


\bibitem[Steyvers et~al\mbox{.}(2022)]%
        {steyvers2022bayesian}
\bibfield{author}{\bibinfo{person}{Mark Steyvers}, \bibinfo{person}{Heliodoro Tejeda}, \bibinfo{person}{Gavin Kerrigan}, {and} \bibinfo{person}{Padhraic Smyth}.} \bibinfo{year}{2022}\natexlab{}.
\newblock \showarticletitle{Bayesian modeling of human--AI complementarity}.
\newblock \bibinfo{journal}{\emph{Proceedings of the National Academy of Sciences}} \bibinfo{volume}{119}, \bibinfo{number}{11} (\bibinfo{year}{2022}), \bibinfo{pages}{e2111547119}.
\newblock


\bibitem[Tversky and Kahneman(1974)]%
        {tversky1974judgment}
\bibfield{author}{\bibinfo{person}{Amos Tversky} {and} \bibinfo{person}{Daniel Kahneman}.} \bibinfo{year}{1974}\natexlab{}.
\newblock \showarticletitle{Judgment under Uncertainty: Heuristics and Biases: Biases in judgments reveal some heuristics of thinking under uncertainty.}
\newblock \bibinfo{journal}{\emph{science}} \bibinfo{volume}{185}, \bibinfo{number}{4157} (\bibinfo{year}{1974}), \bibinfo{pages}{1124--1131}.
\newblock


\bibitem[Wang et~al\mbox{.}(2020)]%
        {wang2020human}
\bibfield{author}{\bibinfo{person}{Dakuo Wang}, \bibinfo{person}{Elizabeth Churchill}, \bibinfo{person}{Pattie Maes}, \bibinfo{person}{Xiangmin Fan}, \bibinfo{person}{Ben Shneiderman}, \bibinfo{person}{Yuanchun Shi}, {and} \bibinfo{person}{Qianying Wang}.} \bibinfo{year}{2020}\natexlab{}.
\newblock \showarticletitle{From human-human collaboration to Human-AI collaboration: Designing AI systems that can work together with people}. In \bibinfo{booktitle}{\emph{Extended abstracts of the 2020 CHI conference on human factors in computing systems}}. \bibinfo{pages}{1--6}.
\newblock


\bibitem[Wang et~al\mbox{.}(2024)]%
        {wang2024scientific}
\bibfield{author}{\bibinfo{person}{Jinge Wang}, \bibinfo{person}{Qing Ye}, \bibinfo{person}{Li Liu}, \bibinfo{person}{Nancy~Lan Guo}, {and} \bibinfo{person}{Gangqing Hu}.} \bibinfo{year}{2024}\natexlab{}.
\newblock \showarticletitle{Scientific figures interpreted by ChatGPT: strengths in plot recognition and limits in color perception}.
\newblock \bibinfo{journal}{\emph{NPJ Precision Oncology}} \bibinfo{volume}{8}, \bibinfo{number}{1} (\bibinfo{year}{2024}), \bibinfo{pages}{84}.
\newblock


\bibitem[Wang et~al\mbox{.}(2023)]%
        {wang2023llm4vis}
\bibfield{author}{\bibinfo{person}{Lei Wang}, \bibinfo{person}{Songheng Zhang}, \bibinfo{person}{Yun Wang}, \bibinfo{person}{Ee-Peng Lim}, {and} \bibinfo{person}{Yong Wang}.} \bibinfo{year}{2023}\natexlab{}.
\newblock \showarticletitle{LLM4Vis: Explainable visualization recommendation using ChatGPT}.
\newblock \bibinfo{journal}{\emph{arXiv preprint arXiv:2310.07652}} (\bibinfo{year}{2023}).
\newblock


\bibitem[Wang et~al\mbox{.}(2022)]%
        {wang2022interpretability}
\bibfield{author}{\bibinfo{person}{Zijie~J Wang}, \bibinfo{person}{Alex Kale}, \bibinfo{person}{Harsha Nori}, \bibinfo{person}{Peter Stella}, \bibinfo{person}{Mark~E Nunnally}, \bibinfo{person}{Duen~Horng Chau}, \bibinfo{person}{Mihaela Vorvoreanu}, \bibinfo{person}{Jennifer Wortman~Vaughan}, {and} \bibinfo{person}{Rich Caruana}.} \bibinfo{year}{2022}\natexlab{}.
\newblock \showarticletitle{Interpretability, then what? editing machine learning models to reflect human knowledge and values}. In \bibinfo{booktitle}{\emph{Proceedings of the 28th ACM SIGKDD Conference on Knowledge Discovery and Data Mining}}. \bibinfo{pages}{4132--4142}.
\newblock


\bibitem[Zhao et~al\mbox{.}(2024)]%
        {zhao2024leva}
\bibfield{author}{\bibinfo{person}{Yuheng Zhao}, \bibinfo{person}{Yixing Zhang}, \bibinfo{person}{Yu Zhang}, \bibinfo{person}{Xinyi Zhao}, \bibinfo{person}{Junjie Wang}, \bibinfo{person}{Zekai Shao}, \bibinfo{person}{Cagatay Turkay}, {and} \bibinfo{person}{Siming Chen}.} \bibinfo{year}{2024}\natexlab{}.
\newblock \showarticletitle{LEVA: Using Large Language Models to Enhance Visual Analytics}.
\newblock \bibinfo{journal}{\emph{IEEE Transactions on Visualization and Computer Graphics}} (\bibinfo{year}{2024}).
\newblock


\end{thebibliography}

\end{document}